\documentclass[preprint,12pt]{elsarticle}
\usepackage{graphicx} 

\usepackage{amssymb}
\usepackage{amsmath}
\usepackage{physics}
\usepackage{xcolor}

\begin{document}

\begin{frontmatter}
\title{A fast and accurate method for simulating Bragg atom interferometers}
\author[berkeley]{Jack Roth}
\ead{jack_roth@berkeley.edu}
\author[berkeley]{Andrew Christensen}
\author[berkeley]{Madeline Bernstein}
\author[berkeley]{Yuno Iwasaki}
\author[berkeley]{Holger Mueller}
\affiliation[berkeley]{
            addressline={Department of Physics, University of California},
            city={Berkeley},
            postcode={94720},
            state={CA},
            country={USA}}
\begin{abstract}
Atom interferometers are used in a variety of applications, from measuring gravity and gravity gradients in the field to performing tests of fundamental physics in the lab. One method of increasing interferometer sensitivity is to produce a larger momentum difference between interferometer arms through the use of large momentum transfer methods, such as Bragg diffraction. However, Bragg diffraction introduces systematic effects in the accumulated interferometer phase that are challenging to characterize.

A Bragg atom interferometer is described by the one-dimensional time-dependent Schrödinger equation (1D-TDSE). In this paper we show that for the case of Bragg diffraction the 1D-TDSE partial differential equation can be separated into several systems of ordinary differential equations, allowing for the use of adaptive step size Runge-Kutta methods. We compare the convergence of this method to the split-step and Crank-Nicolson methods, and present a method for further computational speed-ups using a lookup table.
\end{abstract}
\end{frontmatter}

\section{Introduction}

In atom interferometry, Bragg beamsplitters are used to split atomic wavefunctions into a superposition of momentum states by applying Bragg diffraction. Bragg diffraction works by pulsing a sinusoidal potential on and off in time in such a way that the atom wavefunction scatters off of the potential into a 50-50 superposition of momentum states spaced by an integer number of $h/\ell$, where $\ell$ is the spatial period of the potential. This process is described by the one-dimensional time-dependent Schrödinger equation (1D-TDSE). For atom interferometers making precision measurements it is important that any systematic effects arising from this scattering process are well understood.

One systematic effect of interest stems from the fact that slight differences in initial particle momentum can lead to large differences in the scattering efficiency, and can even cause the wavefunction to scatter into unwanted momentum states. In real-world experiments the atomic wavefunction has non-zero momentum width, and so this effect varies across the wavefunction. These dynamics can be simulated by solving the 1D-TDSE partial differential equation \cite{Fitzek2020} for some initial wavepacket, which is computationally expensive.

In this paper we develop a fast and accurate method for integrating the 1D-TDSE partial differential equation (PDE) for the case of a sinusoidal potential. The method takes advantage of the fact that in momentum space the sinusoidal potential has the form of a delta function, which converts the 1D-TDSE into several systems of ordinary differential equations (ODEs). Thus, the solution to the full PDE problem can be constructed by solving each system of ODEs independently. This method can handle an arbitrary initial wavefunction. For certain problems our method is more efficient than the split-step \cite{Fitzek2020} and Crank-Nicolson \cite{vanDijk2007} methods which have been used to approach this or similar problems previously. The method is compatible with a lookup table, which can be used to reduce computation time. We will derive this simple method, compare its accuracy to the split-step and Crank-Nicolson methods, and provide an example using this method to characterize a systematic effect.

\section{The Method}
\subsection{Derivation}
\label{sec:the_method}

Our goal is to solve for the time evolution of some wavefunction $\ket{\psi}$ under the Schrödinger equation for the Hamiltonian describing Bragg diffraction:
\begin{align}
    \hat{H}=&\frac{\hat{p}^2}{2m}+\hbar\frac{\Omega_\text{eff}(t)}{2}\cos^2(k\hat{x}-\delta t/2), \nonumber \\
    =&\frac{\hat{p}^2}{2m}+\hbar\frac{\Omega_\text{eff}(t)}{8}\qty[\exp(i(2k\hat{x}-\delta t))+\exp(-i(2k\hat{x}-\delta t))+2], \label{eq:hamiltonian_exp}
\end{align}
where $\Omega_\text{eff}(t)$ is a function of time. A derivation of this Hamiltonian can be found in \cite{Kovachy2010}. To solve for the evolution of $\ket{\psi}$ under this Hamiltonian in position space we could employ the split-step or Crank-Nicolson methods (see section \ref{sec:comparison_to_methods}). However, these methods do not take advantage of the fact that in momentum space the potential is a delta function that couples momentum states in discrete intervals of $2\hbar k$.

To take advantage of this property we apply $\bra{p}$ to the left hand side of the Schrödinger equation, noting $\mel{p}{\exp(\pm i2k\hat{x})}{\psi}=\phi(p\mp 2\hbar k)$ and $i\hbar\bra{p}\dv{t}\ket{\psi}=i\hbar\dot{\phi}(p)$ (see \ref{sec:momentum_state_properties} for a derivation of these identities), to yield:
\begin{align} 
    i\hbar\dot{\phi}(p)=&\frac{p^2}{2m}\phi(p) \nonumber \\
    &+\hbar\frac{\Omega_\text{eff}(t)}{8}[\phi(p-2\hbar k)\exp(-i\delta t)+\phi(p+2\hbar k)\exp(i\delta t)+2\phi(p)], \label{eq:pde_momentum_space}
\end{align}
where $\phi(p)$ is the momentum space wavefunction.

Physically, if the potential is formed by an off-resonant AC Stark shift, we can attribute the spacing of $2\hbar k$ between coupled momentum states to the momentum transferred to the atom by the scattering of an off-resonant photon, which is ``absorbed'' and then ``emitted'' by the atom. In the model used here, the electromagnetic field is classical; there are no photons. Since the classical electromagnetic field is described by a sinusoidal potential with a single spatial frequency, momentum states can only be coupled in spacings of this exact frequency.

\begin{figure}
    \centering
    \includegraphics[width=1\linewidth]{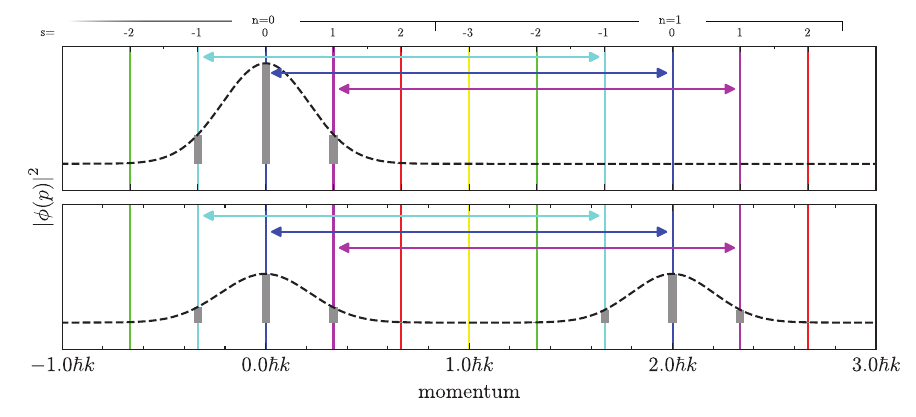}
    \caption{The discretization scheme used when setting $\phi_{n,s}=\phi(2\hbar kn+\hbar k s/S)$ for the case where $S=3$. The upper plot shows an initial momentum space wavefunction, and the lower plot shows the same wavefunction after undergoing a 1st order Bragg pulse. Each discretized point is indicated by a vertical line. For each value of $n$ there is a bin between $n2\hbar k-\hbar k$ and $n2\hbar k+\hbar k$ which contains 6 discretized points. The value of $s$ at each of these points is shown on the top axis. The value of $n$ for each bin is also shown on the top axis. The arrows are each $2\hbar k$ long, and indicate which discretized points are coupled to each other in momentum space (coupled discretized points share the same color). Only three arrows are shown here for simplicity, but in reality every discretized point has two arrows pointing out of it in opposite directions as it is coupled to two other momentum states. Note that coupled states share the same value of $s$, but have different values of $n$. The gray bars indicate the population attributed to each particular discretized state.}
    \label{fig:momentum_discretization}
\end{figure}

To reflect the discrete nature of this coupling (and to make Eq.(\ref{eq:pde_momentum_space}) easier to solve with an ODE solver) we define a discrete set of momentum states $\phi_{n,s}=\phi(2\hbar kn+\hbar k s/S)$. Under this definition the value of $s$ sets how far in momentum space a particular state is from a multiple of $2\hbar k$, and $n$ sets roughly the number of $2\hbar k$ a particular state is from zero momentum. Note that we will enforce $n\in\mathbb{Z}$, $S\in \mathbb{Z},S>0$, and $s\in[-S,-S+1,\dots,S-2,S-1]$. In the limit that $S\to\infty$ this definition for $\phi_{n,s}$ becomes continuous and covers all momenta. A heuristic for choosing $N$ and $S$ is given in section \ref{sec:choice_of_n_s}. A graphical representation of this discretization is provided in Fig.(\ref{fig:momentum_discretization}).

Applying this discretization to Eq.(\ref{eq:pde_momentum_space}) we find
\begin{align}
    i\hbar\dot{\phi}_{n,s}=&\hbar\omega_\text{r}(2 n+s/S)^2\phi_{n,s} \nonumber \\
    &+\hbar\frac{\Omega_\text{eff}(t)}{8}\qty[\phi_{n-1,s}\exp(-i\delta t)+\phi_{n+1,s}\exp(i\delta t)+2\phi_{n,s}], \label{eq:ode_pde_system}
\end{align}
where $\omega_\text{r}=\hbar k^2/2m$ is the recoil frequency. We see that neighboring values of $n$ are coupled together. We also see that for each value of $s$ there is a unique system of ODEs. What would be a PDE in position space has turned into an ODE in momentum space.

Fig.(\ref{fig:momentum_discretization}) reveals that there can be values of $s$ that have little or no amplitude. These values of $s$ can be removed from the calculation, truncating the number of ODEs that must be solved. For initial wavefunctions with a momentum spread much smaller than $2\hbar k$, this can provide a computational speed up in exchange for a loss in accuracy.

To solve for the evolution of an arbitrary initial momentum space wavefunction $\phi(p)$, we simply discretize $\phi(p)$ into $\phi_{n,s}$ and solve the system of ODEs that exists for each value of $s$. A discrete Fourier transform can be used to express a position space wavefunction in the discretized momentum basis:
\begin{align*}
    \phi_j=\frac{1}{L}\sum_{\ell=0}^L\psi_\ell \exp(-i2\pi j\ell/2L),
\end{align*}
where $\psi_\ell=\psi(\ell \delta_x+x_0)$ ($x_0$ is some overall shift), and $L$ is the number of discretized points in position space. The resulting $\phi_j$ maps to $\phi_{n,s}$ via $j=2Sn+s$.

In practice $n$ must be bounded. In the remainder of the text we will assume that the space is truncated such that only $N$ values of $n$ are considered.

\subsection{Choice of $N$ and $S$}
\label{sec:choice_of_n_s}
To solve for the evolution of some state $\ket{\psi}$ (where the position and momentum space wavefunctions are $\psi=\braket{x}{\psi}$ and $\phi=\braket{p}{\psi}$, respectively) we must choose values for $N$ and $S$ such that the calculation has the required spatial extent, momentum extent, spatial resolution, and momentum resolution.

First note that in momentum space the grid spacing is $\hbar k/S$, and the total extent of the simulation is $2\hbar kN$. Therefore, in position space the grid spacing is $1/Nk$, and the total extent of the simulation is $S/k$.

Consider a wavefunction with features at the resolution of $\delta_x$ in position space and $\delta_p$ in momentum space propagating through an atom interferometer with spatial extent $\Delta_x$ and momentum extent $\Delta_p$. For such an interferometer the following inequalities must be met:
\begin{align*}
    \delta_p&\gtrsim\hbar k/S, \\
    \delta_x&\gtrsim 1/Nk, \\
    \Delta_p&\lesssim 2\hbar kN, \\
    \Delta_x&\lesssim S/k.
\end{align*}
Therefore reasonable choices for $N$ and $S$ are $N\gtrsim\max(1/k\delta_x,\Delta_p/2\hbar k)$ and $S\gtrsim\max(\hbar k/\delta_p,\Delta_x k)$.

\section{Comparison to other methods}
\label{sec:comparison_to_methods}

To evaluate the performance of the method presented here we will simulate a Ramsey-Borde interferometer, an interferometer geometry used to measure  recoil frequency. We will compare the performance of this method to the split-step and Crank-Nicolson methods, which can both be used to solve the 1D-TDSE PDE. Both the split-step and Crank-Nicolson methods are designed to handle an arbitrary potential instead of a purely sinusoidal one.

A Ramsey-Borde interferometer accumulates phase $\theta=8Tn_\mathrm{Bragg}^2\omega_\mathrm{r}-Tn_\mathrm{Bragg}\omega_\mathrm{m}$ \cite{Muller2006}, where $T$ is the time between Bragg beamsplitters \cite{Mueller2008, Kovachy2010}, $n_\mathrm{Bragg}$ is the Bragg order, $\omega_\text{r}=\hbar k^2/2m$ is the recoil frequency, and $\omega_\mathrm{m}$ is the amount that the two photon detuning $\delta$ is shifted between the first two and last two Bragg pulses (i.e. the first two Bragg pulses have two-photon detuning $\delta$, the last two Bragg pulses have two-photon detuning $\delta-\omega_\mathrm{m}$).

\begin{figure}
    \centering
    \includegraphics[width=1\linewidth]{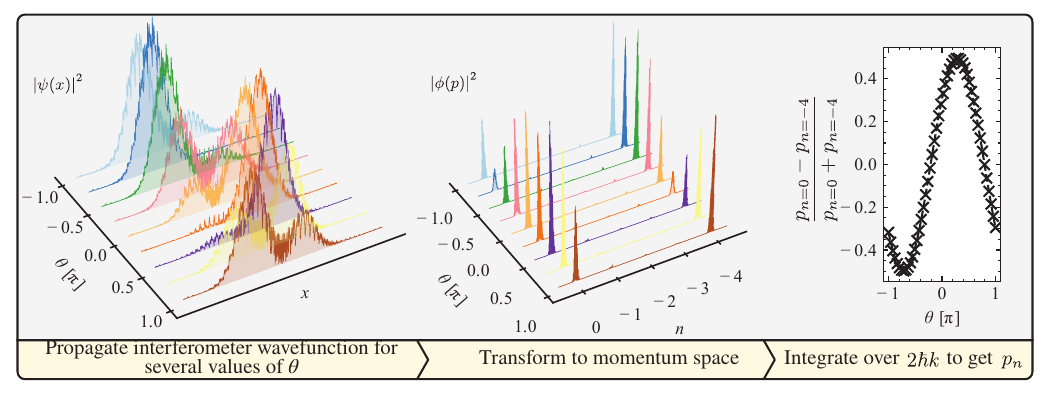}
    \caption{The process by which the split-step, Crank-Nicolson, and the method presented here are compared. Each method is used to simulate a full Ramsey-Borde interferometer composed of Bragg beamsplitters. Simulations are performed for a variety of $\omega_\mathrm{m}$ values, where $\omega_\mathrm{m}$ is set to $\omega_\mathrm{m}=8n_\mathrm{Bragg}\omega_\mathrm{r}-\theta/Tn$ for 50 values of $\theta$ spaced from $-\pi$ to $\pi$. This will trace the interferometer phase over a full fringe. The output wavefunction $\psi(x)$ is then Fourier transformed to obtain the momentum space wavefunction $\phi(p)$, which is integrated over $2\hbar k$ to find $p_n=\abs{\phi_n}^2=\int_{2\hbar k(n-1/2)}^{2\hbar k(n+1/2)}\phi(p')^*\phi(p')\mathrm{d}p$. The ratio of population in each output port is calculated via $(p_{n_\mathrm{initial}}-p_{n_\mathrm{final}})/(p_{n_\mathrm{initial}}+p_{n_\mathrm{final}})$, which is fit to a sinusoid to determine the phase $\theta$ of the simulated interferometer. The difference between this phase and the target phase is due to systematic effects such as parasitic momentum states. These plots were generated with $n_\text{Bragg}=4$ using the method presented here.}
    \label{fig:phase_extraction}
\end{figure}

The simulation is computed for several values of $\omega_\mathrm{m}$, and as a result a fringe is observed in the output populations. The phase $\theta$ of each simulated interferometer is determined by fitting the fringe to a sinusoid. This process is illustrated in Fig.(\ref{fig:phase_extraction}). For more details of the simulated interferometer see \ref{sec:full_ifr_sim_description}.

To simulate this interferometer using the method presented here the DOP853 adaptive step size integrator \cite{2020SciPy,numba} is used to solve Eq.(\ref{eq:ode_pde_system}) with relative and absolute tolerances of $10^{-10}$, a maximum step size of $\Delta t=0.1\omega_\mathrm{r}^{-1}$, $S=60$, and $N=60$. The split-step and Crank-Nicolson methods are used with a spatial step size of $0.018\,k^{-1}$ and several different time step sizes.

\begin{figure}
    \centering
    \includegraphics[width=1\linewidth]{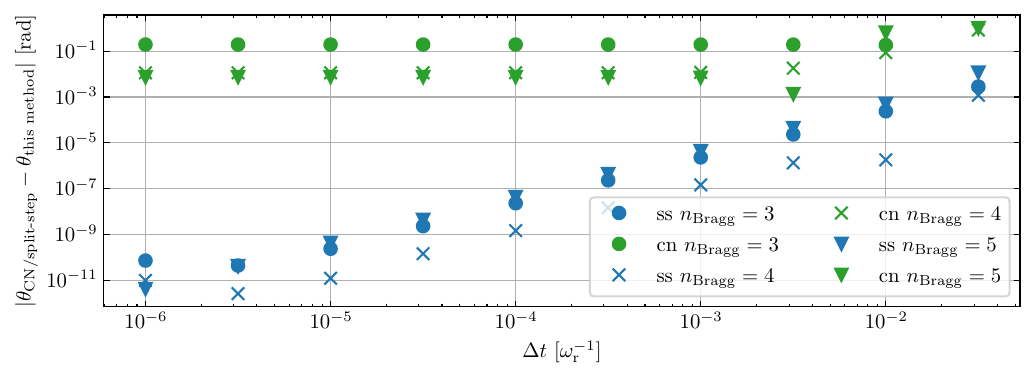}
    \caption{A comparison of the Ramsey-Borde interferometer phase computed by the method presented here, the split-step method (ss), and the Crank-Nicolson method (cn). The interferometer parameters are described in the text.}
    \label{fig:split_step_error}
\end{figure}

Fig.(\ref{fig:split_step_error}) shows the absolute value of the difference between the phase simulated by the method presented here and the split-step and Crank-Nicolson methods. The split-step method converges to the method presented here with a difference of around $10^{-9}\,\mathrm{rad}$ at a step size of approximately $\Delta t=10^{-5}\omega_\mathrm{r}^{-1}$. The method presented here is faster than the split-step method at step sizes below $\Delta t=10^{-3}\omega_\mathrm{r}^{-1}$ on an AMD 3990X. More information about the split-step method can be found in \ref{sec:split_step_derivation}.

From Fig.(\ref{fig:split_step_error}) we see that the Crank-Nicolson method does not converge to our method's result. Decreasing the Crank-Nicolson spatial grid spacing from $0.018k^{-1}$ to $0.011k^{-1}$ increased agreement at step sizes below $\Delta t=3\times10^{-3}\,\omega_\mathrm{r}^{-1}$ to approximately $10^{-3}\,\mathrm{rad}$, which shows that the Crank-Nicolson accuracy is limited by the spatial grid spacing. Further decreases of the spatial grid spacing were limited by available computational resources. From this result it is clear that if a position space solver must be used, the split-step method is preferable to the Crank-Nicolson method. More information about the Crank-Nicolson method can be found in \ref{sec:crank_nicholson_deriv}.

The faster and more accurate performance of our method can be attributed to it's ability to use adaptive step sizes (i.e. larger $\Delta t$ when possible, smaller $\Delta t$ when necessary), and the higher order of the method. While the method presented here outperforms the split-step method at higher levels of precision, it is important to note that the split-step method is able to handle cases where the standing wave deviates from a perfect sinusoid. Such cases are also of importance to atom interferometers, and this difference should be considered when selecting a solver for the case of interest.

 For a thorough comparison of several other 1D-TDSE numerical techniques see \cite{Iitaka1994}.

\section{Investigation of a cloud of atoms in an interferometer}
We now consider a problem that would be intractable with the previously described split-step and Crank-Nicolson methods: simulating a cloud of atoms propagating through an atom interferometer. This problem is important for understanding systematic effects related to parasitic momentum states and the momentum width of the atom wavefunction. To achieve a large speedup over the split-step and Crank-Nicolson methods we will make use of a lookup table as described in \ref{sec:lookup_table_method}. The use of such a lookup table is not possible with the split-step or Crank-Nicholson methods.

\begin{figure}
    \centering
    \includegraphics[width=1\linewidth]{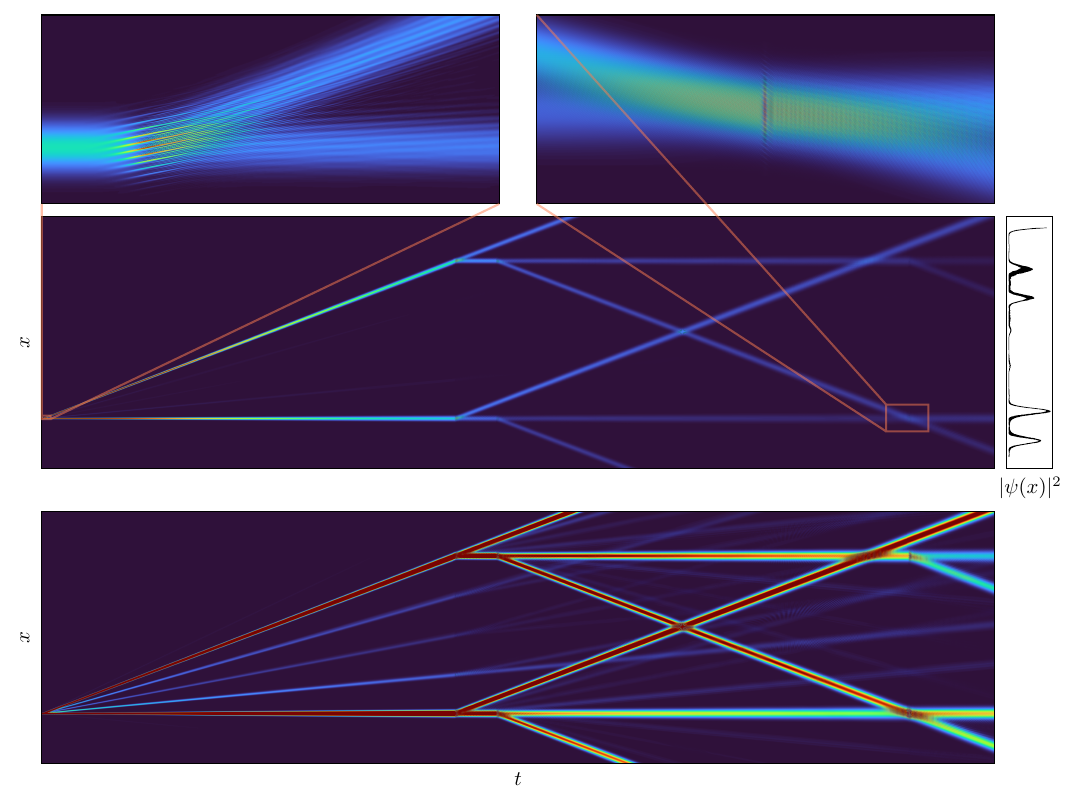}
    \caption{A simulation of a Ramsey-Borde interferometer composed of 4th order Bragg pulses. The displayed heatmap is $|\psi(x,t)|^2$. The Bragg pulses are Gaussian in time with $\sigma=0.26\omega_\text{r}^{-1}$, the time between the first two and last two Bragg pulses is $T=259.62\omega_\text{r}^{-1}$, and the time between the second and third Bragg pulses is $T'=25.96\omega_\text{r}^{-1}$, and the initial wavepacket is a minimum uncertainty Gaussian with $\sigma_p=0.1\hbar k$. In the case of a Cesium atom these numbers correspond to a Bragg pulse with $\sigma=20\,\mathrm{\mu s}$, $T=20\,\mathrm{ms}$, $T'=2\,\mathrm{ms}$ and a velocity spread of a tenth of a recoil velocity. The insets show the first and last Bragg pulses. The bottom plot contains identical data to the middle plot but the maximum color value has been lowered to make the parasitic interferometers clearer. The top right inset displays an area 8 times larger than the top left inset. In this simulation $S=3041$ and $N=15$. Computing the final wavefunction takes $2500\,\mathrm{s}$ on a single core on an AMD 3990X without a lookup table.}
    \label{fig:full_ifr}
\end{figure}

Fig.(\ref{fig:full_ifr}) shows the population as a function of time and position for a full Ramsey-Borde interferometer. In the figure ``parasitic'' momentum states can be seen exiting the imperfect Bragg beamsplitter pulses. These states can bias the population measured in different output momentum states (which shifts the measured phase), or can even interfere during the final interferometer pulse if they spatially overlap with another wavepacket.

As mentioned previously, another systematic effect is related to the momentum width of the atom wavefunction. The resonance condition for the Bragg beamsplitter is only met for a single momentum. Since the wavefunction has width in momentum space, it cannot be that the entire wavefunction is resonant with the Bragg beamsplitter. This can lead to additional systematic effects.

In a real experiment the sinusoidal potential that drives Bragg diffraction is often generated using a standing wave laser beam detuned from resonance with an electric dipole transition. The standing wave laser beam has a finite waist, and so atoms at different locations in the cloud experience different beam intensities and therefore different potential amplitudes. Spatial noise on the beam leads to additional spatially dependent potential amplitudes. Atoms in the cloud also have a distribution of velocities transverse to the standing wave. The velocity distribution of atoms along the standing wave is set by the wavefunction momentum width of the atoms in the cloud. To capture these effects with a simulation, the 1D-TDSE must be solved for a distribution of atom positions and transverse velocities that reflect the distributions in the cloud.

\begin{figure}
    \centering
    \includegraphics[width=1\linewidth]{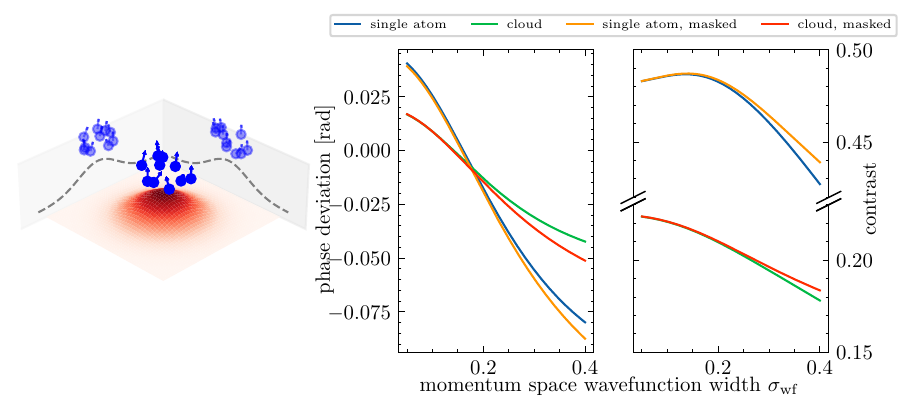}
    \caption{\emph{Left:} This graphic illustrates how atoms in a cloud experience a range of potential depths due to their transverse position spread. Atoms are depicted as solid blue circles. The arrows pointing out of the atoms describe the atom velocity, which can have both vertical and transverse components. The XZ and YZ planes show a projection of the atoms (slightly transparent blue circles), along with arrows depicting their velocity. Propagating the wavefunction of these atoms requires solving the Bragg diffraction Hamiltonian for the potential depth experienced at each point in the atom trajectory. \emph{Right:} These plots show the simulated phase deviation and contrast for a Ramsey-Borde interferometer composed of 4th order Bragg pulses where $\sigma=0.26\omega_\mathrm{r}^{-1}$, $T=64.9\omega_\mathrm{r}^{-1}$, $T'=64.9\omega_\mathrm{r}^{-1}$, $S=1102$, $N=9$, $\Omega_{\mathrm{eff},0}=20\omega_\mathrm{r}^{-1}$, $\sigma_\mathrm{cloud}=1.5\,\mathrm{mm}$, and $\sigma_{v_T,\mathrm{cloud}}=3.5\,\mathrm{mm/s}$. In the case of a Cesium atom these numbers correspond to a Bragg pulse with $\sigma=20\,\mathrm{\mu s}$, $T=5\,\mathrm{ms}$, and $T'=5\,\mathrm{ms}$. 1000 atoms are sampled for the cloud simulations. The phase deviation and contrast curves are generated using the same process described in Fig.(\ref{fig:phase_extraction}). Each simulated interferometer is prepared to zero the accumulated Ramsey-Borde interferometer phase. The meaning of ``masked'' is described in the text. Using a single core on an AMD 3990X it takes $0.6\,\mathrm{s}$ to simulate a full interferometer for each atom in the cloud using the lookup table method. Generating the above plot took 35 hours with multiple cores utilized and required simulating 800,000 single interferometers.}
    \label{fig:full_cloud_ifr_example}
\end{figure}

We present the results from this simulation in Fig.(\ref{fig:full_cloud_ifr_example}). The effect of the momentum space wavefunction width on the measured interferometer phase and contrast is compared for the full cloud of atoms and a single atom. To study the effect of parasitic momentum states on the final interferometer phase, a ``mask'' can be applied in momentum space that removes unwanted momentum states from the simulation after each Bragg pulse. Comparing the ``masked'' and ``unmasked'' interferometer simulations gives insight into how much of the interferometer phase is due to parasitic interferometers.

From these results we immediately see that in all cases there is a dependence on the momentum space wavefunction width at the level of $10$s of milliradians. Existing atom interferometry experiments target this level of precision, and are therefore sensitive to this effect \cite{parker2018,Morel2020,Fan2023}. This effect might also be relevant for atom interferometers aiming to measure gravitational waves \cite{Abe2021}.

Previous studies have investigated the effect of wavefunction momentum width, but have not captured the effects of parasitic interferometers \cite{Estey2015}. The limitations of these studies are discussed in section \ref{sec:spatial_sep_wavepackets}. Split-step and Crank-Nicolson simulations have not been used to simulate a cloud of atoms as these methods are too slow. To the best of our knowledge the effect of wavefunction momentum width in a cloud of atoms with a transverse velocity and with parasitic interferometers has not been studied previously. Our method's enhanced speed allows ensembles of atoms to be simulated in the interferometer, allowing for these effects to be taken into account. This method opens the door to fast simulations of long-baseline Bragg atom interferometry.

\section{Spatial separation of wavepackets}
\label{sec:spatial_sep_wavepackets}

It is informative to consider how the simulation method presented here captures the physics of wavepacket separation. Real Bragg atom interferometers operate in the so-called quasi-Bragg regime \cite{Mueller2008,Bguin2022}. In this regime the momentum width of the wavepackets describing each atom is much less than the bandwidth of the Bragg diffraction process, meaning that the beamsplitter dynamics are well described by a plane wave momentum state. After the Bragg beamsplitter the atom wavepacket is described by a superposition of a diffracted and an undiffracted momentum state. Given enough time these states separate in space and are no longer spatially overlapped. This is evident in Fig.(\ref{fig:full_ifr}), where the insets show that the wavepacket extends over many spatial periods of the Bragg diffraction potential but undergo free evolution for a long enough time to spatially separate.

If we only consider $s=0$, then the simulation describes a true plane wave with infinite spatial extent and therefore no spatial separation. To incorporate the physics of wavepacket separation, it is necessary to include additional values of $s$. For the case of a full Ramsey-Borde interferometer, if wavepacket separation was ignored and wavepackets were approximated as plane waves, interference between different momentum states could be realized after the second beamsplitter pulse. If spatial separation is present this is not possible. To study realistic Bragg interferometers, we must account for this spatial separation.

Previous studies have often represented the wavefunction as a plane wave \cite{Mueller2008,Kovachy2012,Estey2015,Bguin2022}. These studies incorporate spatial separation by tracking the classical motion of each wavepacket between Bragg beamsplitters, and simulating separate Bragg beamsplitter processes for each set of wavepackets that spatially overlap. We refer to this method as the ``hybrid quantum-classical approach.'' The hybrid approach typically does not consider parasitic interferometers, as this requires careful accounting of many classical trajectories.

In the hybrid approach, the effect of non-zero momentum width is studied by repeating the simulation many times with a variety of initial momenta. This is, in effect, what the method presented in this paper is doing: simulating plane wave momentum states for different values of $s$ and then reconstructing the full interferometer output from them. This leads to the perhaps surprising conclusion that the hybrid approaches do not need to manually track the separation of different momentum states if they later average over a distribution of momentum states, as this averaging would capture the effect of spatial separation automatically.

\section{Conclusion}

In summary we have described a method for quickly and accurately propagating the wavefunction of a long-baseline Bragg atom interferometer. This technique is relevant for studying systematic effects in upcoming long-baseline experiments such as gravitational wave detectors \cite{Abe2021} or measurements of the fine-structure constant \cite{parker2018,Morel2020,Fan2023}. The method is computationally cheaper than the split-step and Crank-Nicolson methods. We also described a systematic effect related to the momentum space wavefunction width for a realistic cloud of atoms propagating through an atom interferometer. This effect is relevant for Bragg atom interferometers targeting milliradian precision.

In scenarios where the Bragg pulse intensity profile is fixed in time a lookup table can be used for increased computational efficiency. The primary disadvantage of the method is its inability to treat non-periodic potentials, which are important for understanding certain other systematic effects in atom interferometers. However, the method should find use in quickly simulating systems where the initial wavefunction width or parasitic interferometers are of concern. Future work might investigate how this method can be applied to simulating Bloch oscillations applied to atom interferometers \cite{Clad2017}.

\section*{Acknowledgments}
We acknowledge helpful discussion with Maximilian Beyer (Amsterdam). This work was supported by the U.S. Department of Energy, Office of Science, Office of Basic Energy Sciences [grant number 7709509], Lawrence Berkeley National Laboratory [grant number 7793610], National Science Foundation [grant numbers 2208029, 2512532, and 3068 (the latter administered by Rutgers University)], and the National Aeronautics and Space Administration [grant numbers SUB00001274/URFAO:GR536596 (administered by University of Rochester) and 1720410 (administered by Jet Propulsion Laboratory)].

\bibliographystyle{IEEEtran}
\bibliography{references}

\appendix

\section{The position space approach}

It is simple to recover the PDE that describes the position space wavefunction evolution by applying $\bra{x}$ to the left hand side of the Schrödinger equation where $\hat{H}$ is given by Eq.(\ref{eq:hamiltonian_exp})
\begin{align}
    i\hbar\dot{\psi}(x)=-\frac{\hbar^2}{2m}\pdv[2]{x}\psi(x)+\hbar\frac{\Omega_\text{eff}(t)}{2}\cos^2(kx-\delta t/2)\psi(x) \label{eq:pde_standard}
\end{align}
This PDE can be solved using the split-step method \cite{Fitzek2020} (see \ref{sec:split_step_derivation}) or the Crank-Nicolson method \cite{vanDijk2023,Iitaka1994,vanDijk2007,vanDijk2011,vanDijk2014} (see \ref{sec:crank_nicholson_deriv}).

\section{Ramsey-Borde interferometer parameter selection}
\label{sec:full_ifr_sim_description}

In each simulation discussed in section \ref{sec:comparison_to_methods}, $T=10\omega_\text{r}^{-1}$ and $T'=2\omega_\text{r}^{-1}$. The time-dependent effective Rabi frequency is a truncated Gaussian given by
\begin{align*}
    \Omega_\text{eff}(t,t_0)=\begin{cases}\Omega_{\text{eff},0}\exp(-\frac{(t-t_0)^2}{2\sigma^2}),\quad&\abs{t-t_0}<3\sigma\\0,&\text{otherwise}\end{cases}
\end{align*}
The initial wavefunction is given by
\begin{align*}
    \phi(p)=\frac{1}{\sqrt{\sqrt{2\pi}\sigma_p}}\exp(-\frac{(p/\hbar k-2n_0)^2}{4\sigma_p^2})
\end{align*}
where $\sigma_p=0.14$.

We simulate the full interferometer for three Bragg orders: $n_\text{Bragg}=3$ with $\sigma=0.4966\omega_\text{r}^{-1}$ and $\Omega_{\text{eff},0}=9$; $n_\text{Bragg}=4$ with $\sigma=0.224\omega_\text{r}^{-1}$ and $\Omega_{\text{eff},0}=20$; $n_\text{Bragg}=5$ with $\sigma=0.247\omega_\text{r}^{-1}$ and $\Omega_{\text{eff},0}=30$. These parameters were chosen by selecting the values that produce a 50-50 beamsplitter when integrating Eq.(\ref{eq:ode_pde_system}) where $s=0$.

The two photon detuning $\delta$ is set by the Bragg resonance condition for coupling states $0\hbar k$ and $2n_\mathrm{Bragg}\hbar k$. The resonance condition is $\delta=4n_\mathrm{Bragg}\omega_\mathrm{r}$. This two photon detuning is employed for the first two Bragg pulses in the interferometer. The last two Bragg pulses couple $0\hbar k$ and $-2n_\mathrm{Bragg}\hbar k$ and have resonance condition $-4n_\mathrm{Bragg}\omega_\mathrm{r}=\delta-\omega_\mathrm{m}$ where $\omega_\mathrm{m}=8n_\mathrm{Bragg}\omega_\mathrm{r}$. The resonance condition can be derived from determining the value of $\delta$ that zeros the exponential terms Eq.(\ref{eq:lookup_transformation_eqm}).

\section{Matrix element identities}
\label{sec:momentum_state_properties}

In deriving the following identities we will make use of the following relations
\begin{gather*}
    \braket{x}{\psi}=\psi(x) \qquad \braket{p}{\psi}=\phi(p) \\
    \braket{x}{p}=\exp(ipx/\hbar)/\sqrt{2\pi\hbar} \\
    \mathbb{I}=\int_{-\infty}^\infty\dyad{x'}\dd{x'}=\int_{-\infty}^\infty\dyad{p'}\dd{p'}
\end{gather*}
where $\ket{p}$ denotes a plane wave with momentum $p$.

Starting with the time-derivative term
\begin{align*}
    i\hbar\bra{p}\dv{t}\ket{\psi}=&i\hbar\bra{p}\dv{t}\mathbb{I}\ket{\psi} \\
    =&i\hbar\dv{t}\int_{-\infty}^\infty\braket{p}{p'}\braket{p'}{\psi}\dd{p'}=i\hbar\dot{\phi}(p)
\end{align*}
The $\exp(\pm i2k\hat{x})$ type terms evaluate to
\begin{align}
    \mel{p}{\exp(\pm i2k\hat{x})}{\psi}=&\mel{p}{\exp(\pm i2k\hat{x})\mathbb{I}}{\psi} \nonumber \\
    =&\int_{-\infty}^\infty \mel{p}{e^{\pm i2k\hat{x}}}{p'}\braket{p'}{\psi}\dd{p'} \nonumber \\
    =&\int_{-\infty}^\infty \delta(p'-p\pm 2\hbar k)\phi(p')\dd{p'} \nonumber \\
    =&\phi(p\mp 2\hbar k) \label{eq:mel_proof}
\end{align}
where in the second to last step we used the following relation
\begin{align}
    \mel{p}{\exp(\pm i2k\hat{x})}{p'}=&\mel{p}{\mathbb{I}\exp(\pm i2k\hat{x})\mathbb{I}}{p'} \nonumber \\
    =&\int_{-\infty}^\infty\int_{-\infty}^\infty\braket{p}{z}\mel{z}{\exp(\pm i2k\hat{x})}{z'}\braket{z'}{p'}\dd{z'}\dd{z} \nonumber \\
    =&\frac{1}{2\pi\hbar}\int_{-\infty}^\infty\int_{-\infty}^\infty \exp(-ipz/\hbar)\braket{z}{z'}\exp(\pm i2kz)\exp(ip'z'/\hbar)\dd{z'}\dd{z} \nonumber \\
    =&\frac{1}{2\pi\hbar}\int_{-\infty}^\infty \exp(i\qty(p'-p\pm 2\hbar k)z/\hbar)\dd{z} \nonumber \\
    =&\delta(p'-p\pm 2\hbar k) \label{eq:mel_exp_deltafnc}
\end{align}

\section{Lookup table transformation}
\label{sec:lookup_table_method}
Consider the frame transformation $\tilde{\phi}_{n,s}=\exp(i(2n+s/S)^2\omega_\text{r}t)\phi_{n,s}$ applied to Eq.(\ref{eq:ode_pde_system})
\begin{align}
    i\hbar\dot{\tilde{\phi}}_{n,s}=&\hbar\frac{\Omega_\text{eff}(t)}{8}\left[\tilde{\phi}_{n-1,s}\exp(i(-4-8n+4s/S)\omega_\text{r}t)\exp(-i\delta t)\right. \nonumber \\
    &\left.+\tilde{\phi}_{n+1,s}\exp(i(-4-8n-4s/S)\omega_\text{r}t)\exp(i\delta t)+2\tilde{\phi}_{n,s}\right] \nonumber \\
    =&\hbar\frac{\Omega_\text{eff}(t)}{8}\left[\tilde{\phi}_{n-1,s}\exp(i(-4-8n)\omega_\text{r}t)\exp(-i(\delta-4s\omega_\text{r}/S) t)\right. \nonumber \\
    &\left.+\tilde{\phi}_{n+1,s}\exp(i(-4-8n)\omega_\text{r}t)\exp(i(\delta-4s\omega_\text{r}/S) t)+2\tilde{\phi}_{n,s}\right] \label{eq:lookup_transformation_eqm}
\end{align}
In this frame the momentum shift characterized by $s$ now appears as a modification to the two-photon detuning instead of a kinetic energy term. When simulating a cloud of atoms propagating through a matterwave interferometer this frame is particularly useful as a lookup table can be used to evaluate solutions to $i\hbar\tilde{\phi}_{n,s}$, as shifts in $\delta$ are equivalent to the shifts in $s$.

For a fixed pulse envelope $\Omega_\text{eff}(t)$ we make a lookup table for the evolution of $\phi_{n=0,s=0}$ as a function of $\delta$ and the amplitude of $\Omega_\mathrm{eff}(t)$. By shifting $\delta$ by the appropriate amount we can recover the evolution of $\phi_{n=0,s}$ for any $s$. Further frame transformations allow us to recover $\phi_{n,s}$ for any $n$ and any $s$. This provides a speed up that makes it computationally tractable to simulate an ensemble of atoms propagating through the interferometer with standard hardware.

It is straightforward to generate a lookup table that produces agreement at the $10^{-7}$ level in population with direct integration of Eq.(\ref{eq:ode_pde_system}).

\section{The split-step method}
\label{sec:split_step_derivation}

The split-step method relies on the approximation
\begin{align}
    \exp(-\frac{i}{\hbar}\hat{H}\Delta t)\approx\exp(-\frac{iH_x\Delta t}{2\hbar})\exp(-\frac{iH_p\Delta t}{\hbar})\exp(-\frac{iH_x\Delta t}{2\hbar}) \label{eq:split_step_equation}
\end{align}
The first exponential term is applied to the position space wavefunction, and the result is fast Fourier transformed (FFT) to obtain the momentum space wavefunction. The middle exponential term is then applied to the momentum space wavefunction, after which an inverse FFT is performed on the result to obtain the position space wavefunction again before the final exponential term is applied.

From Glauber's theorem we note
\begin{align}
\exp(A)\exp(B)\exp(A)=\exp(2A+B+[[A,B],A]/4) \label{eq:split_step_derivation_glauber}
\end{align}
Consider some Hamiltonian $H=H_p+H_x$. Typically $H_p$ is easier to propagate in momentum space, and $H_x$ is easier to propagate in position space. For some small step in time the state $\ket{\psi}$ will evolve under this Hamiltonian via 
\begin{align*}
    \ket{\psi(t+\delta t)}=\exp(-\frac{i}{\hbar}\hat{H}\delta t)\ket{\psi(t)}
\end{align*}
Using Eq.(\ref{eq:split_step_derivation_glauber}) we can expand the exponential. Setting $A=-iH_x\delta t/2\hbar$ and $B=-iH_p\delta t/\hbar$ we find
\begin{align*}
    \exp(-\frac{i}{\hbar}\hat{H}\delta t+\frac{i}{16\hbar^3}\comm{\comm{H_x}{H_p}}{H_x}\delta t^3)=\exp(-\frac{iH_x\delta t}{2\hbar})\exp(-\frac{iH_p\delta t}{\hbar})\exp(-\frac{iH_x\delta t}{2\hbar})
\end{align*}
For small $\delta t$ the $\delta t^3$ term is negligable, and can be ignored. This leads to the split split-step approximation in Eq.(\ref{eq:split_step_equation}). We implement the split-step method in Python using the Numba JIT compiler \cite{numba} to produce performant code.

\section{The Crank-Nicolson method}
\label{sec:crank_nicholson_deriv}

The Crank-Nicolson method for the one-dimensional time-dependent Schrödinger equation can be derived by approximating the unitary evolution operator via the 1,1 order Pad\'e approximant \cite{Iitaka1994,vanDijk2011,vanDijk2014}. Applying this to some initial wavefunction $\psi(x,t)$ we find
\begin{align*}
    \psi(x,t+\Delta t)=U(\Delta t)\psi(x,t)\approx\frac{1-iH\Delta t/2\hbar}{1+iH\Delta t/2\hbar}\psi(x,t)=B^{-1}A\psi_j^k
\end{align*}
 where in the last step we discretized $\psi(x,t)$ via $\psi^k_j=\psi(j\Delta x,k\Delta t)$. Using a second order central finite difference to treat the $\pdv[2]{x}$ in the momentum term in the Hamiltonian will result in $A$ and $B$ being tridiagonal matrices with components as follows 
\begin{align}
    A_{j,j}=&\qty[1-iH^k\Delta t/2\hbar]_{j,j}=1-i\frac{\Delta t}{2}\qty(V^k_{j}+\frac{1}{\Delta x^2}) \\
    B_{j,j}=&\qty[1+iH^k\Delta t/2\hbar]_{j,j}=1+i\frac{\Delta t}{2}\qty(V^k_{j}+\frac{1}{\Delta x^2}) \\
    A_{j,j-1}=&A_{j-1,j}=\qty[1-iH^k\Delta t/2\hbar]_{j,j-1}=i\frac{\Delta t}{4\Delta x^2} \\
    B_{j,j-1}=&B_{j-1,j}=\qty[1+iH^k\Delta t/2\hbar]_{j,j-1}=-i\frac{\Delta t}{4\Delta x^2}
\end{align}
where $H^k=H(k\Delta t)$ and $V^k=V(k\Delta t)$. For a discussion of higher-order Crank-Nicolson type routines see \cite{vanDijk2023,vanDijk2007}. A single timestep can be executed in linear time through the use of the Thomas tridiagonal matrix division algorithm.

We implement the Crank-Nicolson method in Python using the Numba JIT compiler \cite{numba} to produce performant code. The Python Crank-Nicolson time step function was found to have a comparable runtime (within a factor of 2) of an equivalent Fortran and an equivalent C implementation.

\end{document}